# Development of Sapphire-Based Solid-State Neutron Detectors for Fusion Environments

Matteo Hakeem Kushoro[a,b*], Stephanie Cancelli[c], Gabriele Croci[a], Benno Margesin[c], Federica Mantegazzini[c], and Enrico Perelli Cippo[b]

a. Department of Physics "Giuseppe Occhialini", University of Milano-Bicocca, Piazza dell'Ateneo Nuovo 1, 20126, Milano, Italy
b. Institute for Plasma Science and Technology - CNR, Via Roberto Cozzi 53 20125, Milano, Italy
c. Fondazione Bruno Kessler (FBK), Via Sommarive 18, Trento, Italy

* Corresponding Author (matteo.kushoro@unimib.it)

## Abstract

Single-crystal diamond is the reference solid-state technology for fast neutron detection in fusion devices. Despite this, its complex manufacturing increases their costs and puts limitations on their geometry. Moreover, the limits in their long-term stability and their behaviour in harsh environments further constraint its applications. Synthetic sapphire ($Al_2O_3$) is a promising alternative as a solid state detector material: it is available in large, low-cost wafers and features both a wider bandgap and a higher displacement energy compared to diamond. Despite this, it has so far received little-to-no attention as a neutron detector. This work presents the first steps of a development programme for sapphire-based fast neutron detectors. Eight prototypes with 4.5 × 4.5 mm² and 9 × 9 mm² active areas and 625 μm thickness were fabricated from commercial wafers with Al-1%Si electrodes and coupled to different electronic configurations. All prototypes detected 5.47 MeV alpha particles from a $^{241}$Am source, confirming the viability of both the fabrication process and the use of sapphire as an active material. The small signals (30–40 mV over a peak-to-peak noise of about 8 mV) and the count rate increasing with bias indicate that sapphires will require higher electric fields than the ones experimented in the work (up to ≃2.5 kV). Geant4 simulations were also used to estimate the response of a sapphire detector to 14.06 MeV (DT) neutrons, highlighting the perspectives of the detector as a fusion diagnostic.

**Keywords:** Sapphire; Solid State Detectors; Fast Neutrons; Neutronics; Tokamak Diagnostics

## 1. Introduction

Fast neutron detection plays a central role in many nuclear facilities and experiments, such as fusion plasma devices, accelerator facilities and spallation neutron sources. In fusion devices, fast neutron counters and spectrometers are needed to assess key plasma parameters - such as fusion power, fuel ion ratio and ion temperature (1). Solid-state detectors (SSDs) are well suited to these tasks thanks to their compactness, mechanical robustness, radiation hardness and low sensitivity to gamma radiation (2) (3) (4).

Among SSDs, single-crystal diamond (SCD) has become the reference technology for fast neutron spectroscopy. Its wide bandgap, high carrier mobility and low leakage current (5) (6) (7) result in very low noise, excellent energy resolution (8) and fast signal response (9), while their small size and intrinsic gamma discrimination (10) give them a clear advantage over scintillator-based systems. These properties have been demonstrated in major fusion experiments such as JET, where diamond detectors provided real-time, high-rate neutron spectroscopy (11) (12), allowing for the measurement of ion temperature and fuel ion ratio (13) (14). Outside of fusion, diamond detectors are also used in nuclear facilities beam monitoring (15), dosimetry (5), aerospace (16) and industrial applications (17).

Despite these results, diamond technology still has limitations that hinder its wider adoption. Electronic-grade SCD is only available in small sizes (typically few millimetres across) and costs several thousand euros per crystal (18). This makes it prohibitive to build large-area detectors, arrays or pixelated devices, which are increasingly needed for spatially resolved measurements. In addition, past experiences on SCDs revealed that their charge collection efficiency degrades progressively under prolonged exposure to high neutron fluences (19) much faster than other SSDs (20) and that their behaviour degrades in environments characterized by temperatures higher than 230°C (2). These factors put limitations in the use of SCDs for next-generation fusion devices, characterized by high neutron fluxes and harsh environments: as such, there is motivation to the search for other wide-bandgap materials that retain the advantages of diamond without the same practical constraints.

Synthetic sapphire (crystalline $Al_2O_3$) is a promising candidate. Since its manufacturing technology is well established in the optics industry, high-quality wafers are widely available at low cost (on the order of tens €/cm$^2$) and in much larger sizes than synthetic diamond (21). Compared with diamond, sapphire has a comparable resistivity (about $10^{16}$ Ω·cm), a higher displacement energy (79 eV vs 43 eV) and a wider bandgap (9.9 eV vs 5.5 eV) (22) (23). While such a bandgap is usually a drawback for conventional electronic devices, in high-radiation and high-temperature environments it becomes an asset since it points to extremely low leakage currents, good radiation hardness, and an intrinsic discrimination against lower-energy radiation components. Despite this, sapphire has received limited attention as an active material for solid state neutron detectors, with the few existing studies (21) (22) (24) (25) being mainly proof-of-concept demonstrations addressing applications outside neutron diagnostics. Its neutron detection performance, spectroscopic capabilities, long-term stability, and response under high neutron fluence and at high temperature remain therefore largely unexplored. Moreover, no detector development programme has yet studied sapphire as a scalable SSD technology, from material selection and detector fabrication to characterization and application-oriented validation.

This article presents the first steps of such a programme. We first use Geant4 simulations to estimate the response and the expected detection efficiency of a sapphire detector for 14 MeV neutrons, representative of the products of D-T fusion reactions (section 2). We then describe the fabrication of the first prototypes, from commercial synthetic sapphire substrates through cleaning, metallization, dicing, contacting and casing, together with the development of a dedicated readout PCB and the selection of an appropriate electronic chain (section 3). The prototypes are then characterized with an alpha source, studying their response as a function of the applied bias voltage (section 4). Finally, we discuss the current limitations of the approach and the activities planned towards the use of sapphire-based SSDs in fusion devices and other harsh environments (section 5).

## 2. Nuclear Reaction Channels and Geant4 Simulation

The response of SSDs to fast neutron irradiation depends on the multiple nuclear reaction channels through which the neutron is converted to the charged particles that allow the signal generation (26). As such, the response is usually complex, relying on the different cross-sections and Q-values of the reaction channels as well as on the behaviour of the generated charged particles (which might be able to escape the active volume before depositing all the energy) (27) (28). To study this, we used the Geant4 toolkit to simulate the response of a sapphire detector under 14.06 MeV neutrons (which are the radiation of most interest for nuclear fusion technologies). Geant4 simulates the interaction of the neutron with matter, tracking both the path and the energy deposition of the secondary particles within the active volume and the surrounding elements. As such, it is able to take into account partial deposition effects (like the ones caused by secondary particles escaping the active volume before depositing their entire energy) and secondary particles generated by other non-active elements of the detector, like the casing or the printed circuit board (PCB) support.

The detector was modelled having an active volume of 9mm x 9mm x 625µm (which is representative of the prototypes described in section 3) and was irradiated with $4 \cdot 10^7$ simulated neutrons fired in a 9mm x 9mm parallel beam. To simulate the effect of the finite energy resolution, a 280keV gaussian broadening was applied to all $E_d$ bins: such value was chosen to emulate the energy resolution obtained with other prototypical SSDs, such as the Silicon Carbide (2% energy resolution for DT neutrons) (29).

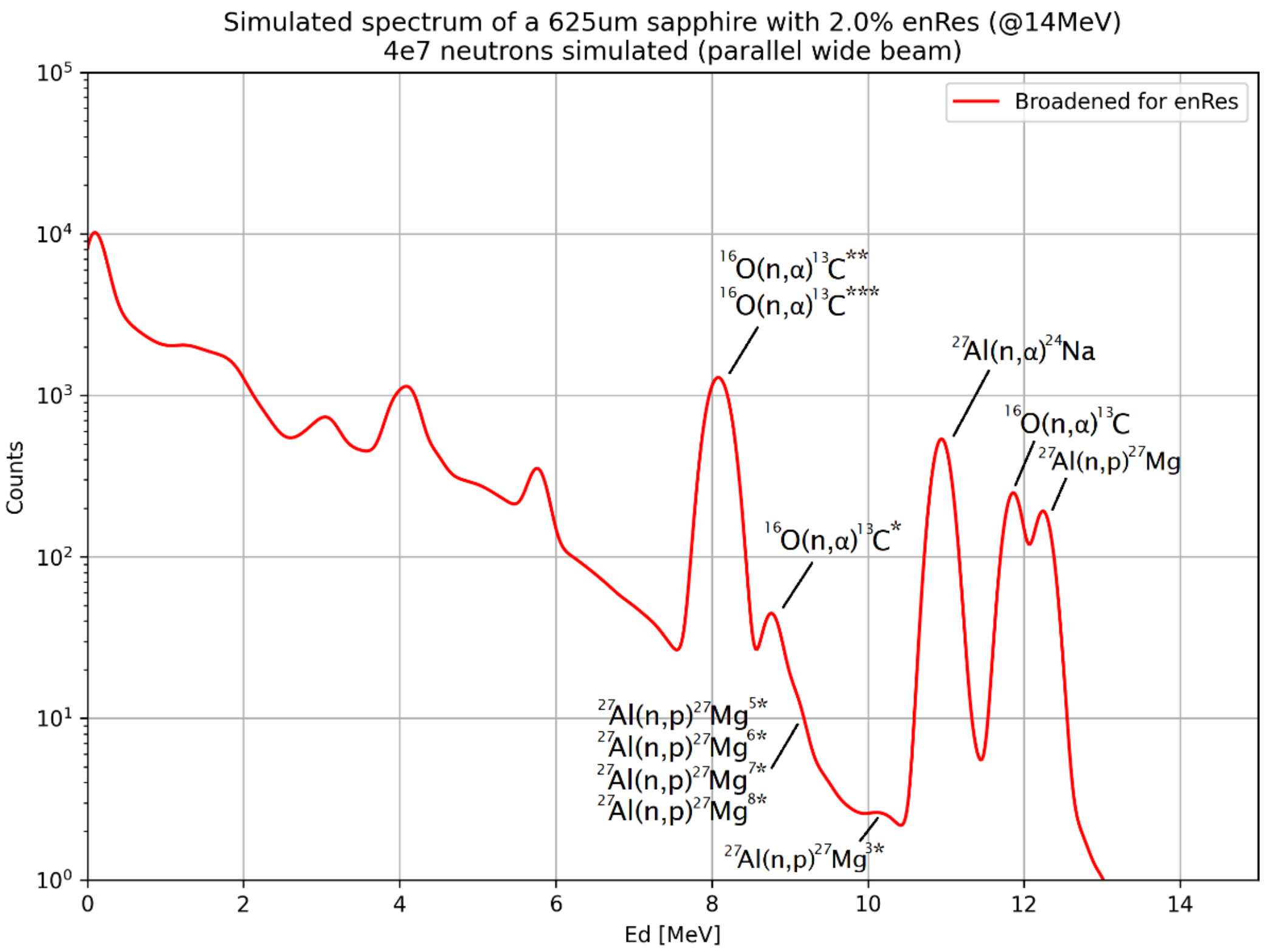

*Figure 1: Pulse height spectrum of a 625μm sapphire detector under DT neutrons obtained with Geant4. The spectrum includes a gaussian broadening that accounts for a supposed 2% energy resolution for the 14MeV neutrons. The main reactions producing an $E_d$ > 8 MeV are reported in figure, highlighting the respective spectral feature.*

Figure 1 shows the results of the simulation for DT neutrons in the form of a deposited energy ($E_d$) spectrum. The spectrum features numerous peaks that are due to the different reaction channels that neutron undergo with a nucleus of the active volume. The three main reaction channels are due to the reactions $^{27}Al(n,p)^{27}Mg$ and $^{27}Al(n,\alpha)^{24}Na$ with aluminium and the $^{16}O(n,\alpha)^{13}C$ with oxygen. These result in charge particles (p, α, and ionized $^{27}Mg$, $^{24}Na$ or $^{13}C$) having a kinetic energy $E_d$ equal to the neutron energy minus the Q-value of the reaction, which is different for each. Moreover, the reaction might cause the resulting nucleus to be in a nuclear excited state (through a process that is usually referred to as "inelastic reaction"), which decays through isometric transition producing gammas that escape the active volume: as such, the energy of the excited level does not produce e-h pairs in the active volume, and is therefore subtracted from $E_d$. Table 1 reports all the reactions (elastic and anelastic) whose resulting $E_d$ is larger than 7.99 MeV, which was used as a guide to identify the features in Figure 1. Note that, while the resulting $^{27}Mg$ and $^{24}Na$ are unstable isotopes that experience beta decay, their decay times (9.46 minutes (30) and 14.96 hours (31), respectively) are much longer than the typical charge collection time of the electronics (≃nanoseconds); as such, such processes are of no interest for the detection process, since they don't happen simultaneously with the charge collection that creates the detected signals.

| Reaction | Q-Value (32) | Nuclear Excited State Energy (30) (31) (33) | Resulting $E_d$ (full deposition) |
|---|---|---|---|
| $^{27}Al(n,p)^{27}Mg$ | 1.83 MeV | 0 | 12.23 MeV |
| $^{27}Al(n,p)^{27}Mg^*$ | 1.83 MeV | 0.98 MeV | 11.25 MeV |
| $^{27}Al(n,p)^{27}Mg^{2*}$ | 1.83 MeV | 1.70 MeV | 10.53 MeV |
| $^{27}Al(n,p)^{27}Mg^{3*}$ | 1.83 MeV | 1.94 MeV | 10.29 MeV |
| $^{27}Al(n,p)^{27}Mg^{4*}$ | 1.83 MeV | 3.11 MeV | 9.12 MeV |
| $^{27}Al(n,p)^{27}Mg^{5*}$ | 1.83 MeV | 3.43 MeV | 8.80 MeV |
| $^{27}Al(n,p)^{27}Mg^{6*}$ | 1.83 MeV | 3.48 MeV | 8.75 MeV |
| $^{27}Al(n,p)^{27}Mg^{7*}$ | 1.83 MeV | 3.49 MeV | 8.74 MeV |
| $^{27}Al(n,p)^{27}Mg^{8*}$ | 1.83 MeV | 3.56 MeV | 8.69 MeV |
| $^{27}Al(n,p)^{27}Mg^{9*}$ | 1.83 MeV | 3.76 MeV | 8.49 MeV |
| $^{27}Al(n,p)^{27}Mg^{10*}$ | 1.83 MeV | 3.79 MeV | 8.46 MeV |
| $^{27}Al(n,p)^{27}Mg^{11*}$ | 1.83 MeV | 3.88 MeV | 8.35 MeV |
| $^{27}Al(n,p)^{27}Mg^{12*}$ | 1.83 MeV | 4.15 MeV | 8.08 MeV |
| $^{16}O(n,\alpha)^{13}C$ | 2.22 MeV | 0 | 11.84 MeV |
| $^{16}O(n,\alpha)^{13}C^*$ | 2.22 MeV | 3.09 MeV | 8.75 MeV |
| $^{16}O(n,\alpha)^{13}C^{2*}$ | 2.22 MeV | 3.68 MeV | 8.16 MeV |
| $^{16}O(n,\alpha)^{13}C^{3*}$ | 2.22 MeV | 3.85 MeV | 7.99 MeV |
| $^{27}Al(n,\alpha)^{24}Na$ | 3.13 MeV | 0 | 10.93 MeV |
| $^{27}Al(n,\alpha)^{24}Na^*$ | 3.13 MeV | 0.47 MeV | 10.46 MeV |
| $^{27}Al(n,\alpha)^{24}Na^{2*}$ | 3.13 MeV | 0.56 MeV | 10.37 MeV |
| $^{27}Al(n,\alpha)^{24}Na^{3*}$ | 3.13 MeV | 1.34 MeV | 9.59 MeV |
| $^{27}Al(n,\alpha)^{24}Na^{4*}$ | 3.13 MeV | 1.34 MeV | 9.59 MeV |
| $^{27}Al(n,\alpha)^{24}Na^{5*}$ | 3.13 MeV | 1.35 MeV | 9.58 MeV |
| $^{27}Al(n,\alpha)^{24}Na^{6*}$ | 3.13 MeV | 1.51 MeV | 9.42 MeV |
| $^{27}Al(n,\alpha)^{24}Na^{7*}$ | 3.13 MeV | 1.85 MeV | 9.08 MeV |
| $^{27}Al(n,\alpha)^{24}Na^{8*}$ | 3.13 MeV | 1.88 MeV | 9.05 MeV |
| $^{27}Al(n,\alpha)^{24}Na^{9*}$ | 3.13 MeV | 2.51 MeV | 8.42 MeV |
| $^{27}Al(n,\alpha)^{24}Na^{10*}$ | 3.13 MeV | 2.56 MeV | 8.37 MeV |

*Table 1: List of the possible nuclear reactions between a DT neutron and a nucleus contained in the sapphire. Reactions are listed with the relative $E_d$, obtained as the neutron energy (14.06 MeV) minus the reaction Q-value and the excited nuclear state level of the product. Only reactions with $E_d \geq 7.99$ MeV are included.*

The spectrum resulting from this (see Figure 1) feature multiple peaks that partially overlap. The most distinguishing features are those due to (n,α) reactions: this is possibly due to the fact that protons from (n,p) reactions, being lighter, get a larger fraction of the total kinetic energy and are less easily stopped by the sapphire material; as such, protons have a much higher probability to escape and not deposit the entirety of the $E_d$ into the detector. The four reactions $^{27}Al(n,\alpha)^{24}Na$, $^{16}O(n,\alpha)^{13}C$, $^{16}O(n,\alpha)^{13}C^{2*}$, and $^{16}O(n,\alpha)^{13}C^{3*}$ are the ones with the largest cross sections to 14 MeV neutrons (7 mBarn, 2 mBarn, 5 mBarn, and 6 mBarn, respectively (34)) and are therefore projected to be the main features to be exploited for a future detector.

## 3. Detector Prototypes and Manufacturing

The prototypes were designed as simple parallel plate capacitors using a common layout editor. The design includes a total of three layers, two for the plates of the capacitor (top and bottom) and a third layer for contact holes (for the passivation of the front electrode with a dielectric layer in future iterations).

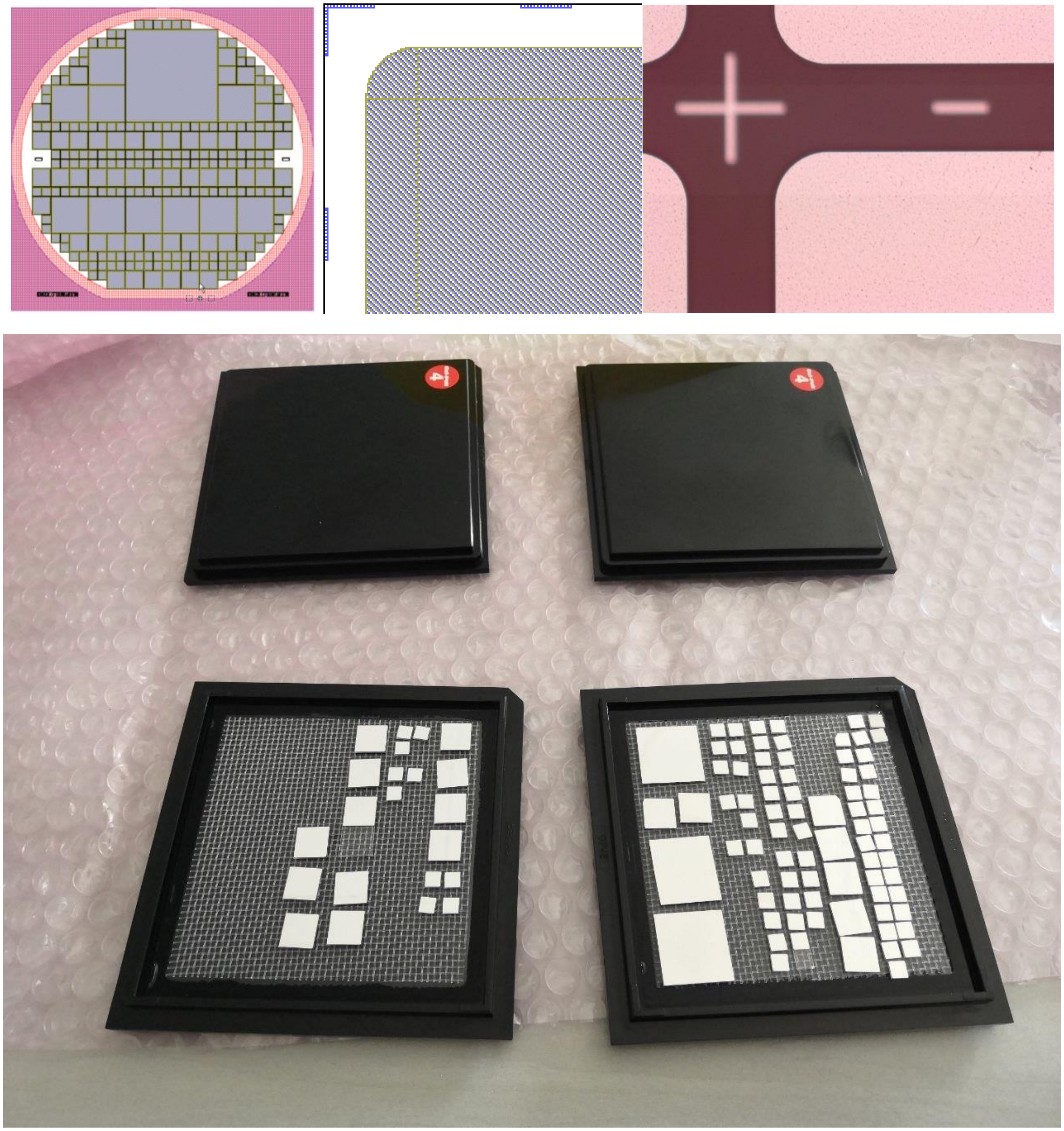

*Figure 2: Left: Sapphire wafer layout with the four different device sizes. Centre: Detail of the rounded corners of the parallel plate capacitor. Right: Image of the corner of a bottom electrode after dry etching. Bottom: Picture of the manufactured sapphire active volumes.*

The active volume was manufactured from synthetic sapphire wafers produced by SITUS Technicals GmbH. The SEMI standard 6′′ wafers with a thickness of 625µm allowed for a large degree in customization in the size of the detectors' active volumes. The fabrication process was conducted in the cleanroom of Fondazione Bruno Kessler (FBK), preceded by a cleaning of the wafers using the standard RCA procedure (sulfuric acid, ammonia, hydrofluoric acid and chloride acid baths). Both faces of the sapphire wafer were metallized with 500 nm thick aluminium containing 1% silicon (Al-1%Si). The aluminium films were then patterned by photolithography and etched with a dry etching process (first on the front side and then on the backside), producing a parallel-plate capacitor covering the whole die, as shown in Figure 2. The electrode edges were kept 200 µm from the die edge and their corners rounded to a 300 µm radius to limit discharges at high voltage.

Finally, the wafer was diced with a 100 µm blade into square devices with nominal sides of 4.5mm, 9mm, 18mm or 36mm (as presented in Figure 2). The first size was chosen to mimic the typical dimension of SCDs, while the others were chosen as a testbed for investigating the functionality of larger area devices. Only 4.5mm and 9mm were used for the scope of this article. Since the actual kerf is 120–130 µm, the real sides are 20–30 µm smaller than nominal, with a 10–20 µm uncertainty due to dicing blade wear. A picture of the completed active volumes is reported in Figure 2.

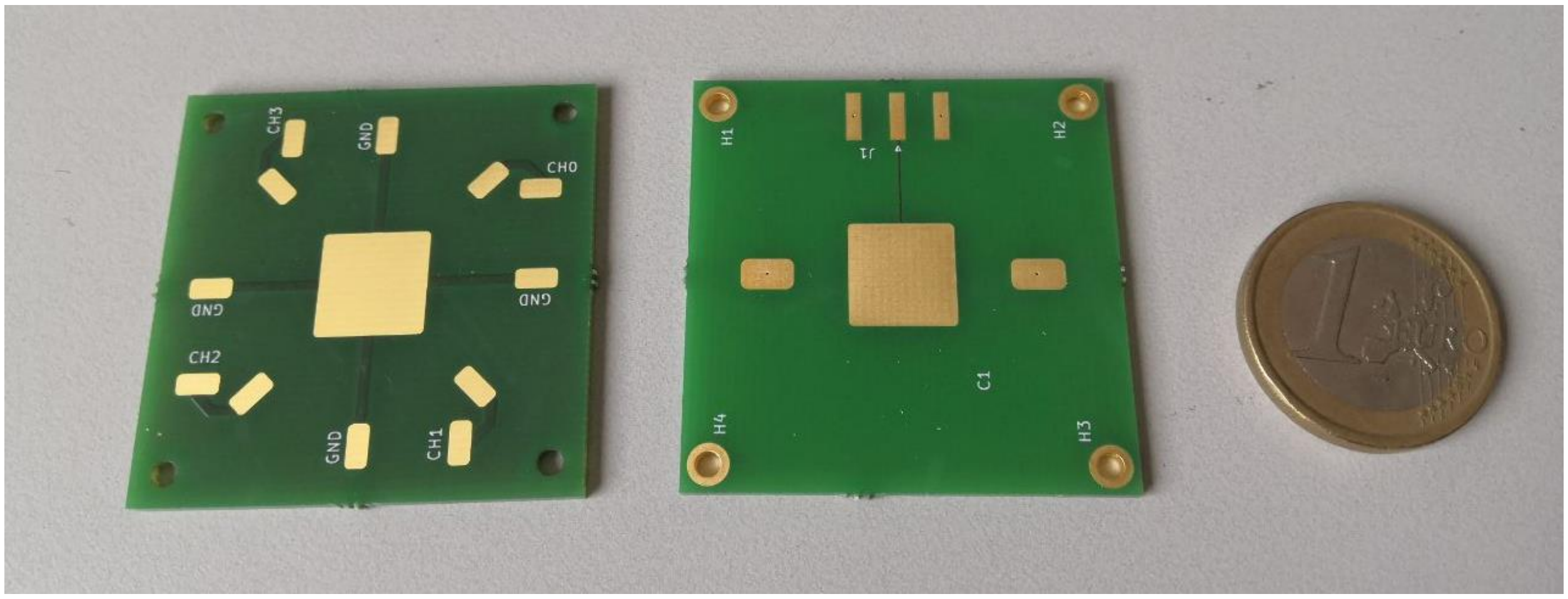

*Figure 3: Bottom: Picture of the Type 1 (left) and Type 2 (right) designed PCB for the sapphire detector. A coin is included for size reference.*

In parallel, two custom Printed Circuit Boards (PCBs) were designed in the Institute of Plasma Science and Technology of Milan (ISTP-CNR) to serve as a support for the active volume, as shown in Figure 3. The goal of the designs was to minimize the probability of the high levels of voltage bias inducing discharges on the surface or in. The first design ("Type 1") was manufactured to support a SHV connector (to accommodate high voltage bias requirements within the safety margin of the connector), while the second design ("Type 2") was manufactured to support a SMA connector (which is the one used for most SSDs). A third option was a PCB which is routinely used for diamond detectors, supporting a SMA connector. A fourth "sandwich" option was instead made of two boards encasing the active volume, with the contact being achieved through a hole at the centre of the two boards. Sapphire active volumes were installed on the PCBs by the means of a conductive glue. An insulating paint was then applied on the exposed PCB components to further reduce the risk of discharges. Lastly, the prototypes were installed in aluminium boxes to provide both structural stability and insulation from EM noise.

Eight prototypes were manufactured to experiment different connectors, casings, and active volume sizes, as well as to widen the experimental sample so that the observed response could be verified as reproducible across independent devices rather than relying on a single detector. A list of the prototypes (with relative characteristics) is reported in Table 2. A picture of the finished prototypes is reported in Figure 4.

| Prototype Name | Active Volume Size | Connector Type | PCB type |
| --- | --- | --- | --- |
| Sapphire-A | 4.5mm x 4.5mm x 625 µm | SMA | Diamond-type |
| Sapphire-B | 4.5mm x 4.5mm x 625 µm | SHV | Diamond-type |
| Sapphire-C | 4.5mm x 4.5mm x 625 µm | SHV | Diamond-type |
| Sapphire-D | 9mm x 9mm x 625 µm | SMA | Type 2 |
| Sapphire-E | 9mm x 9mm x 625 µm | SHV | Type 1 |
| Sapphire-F | 9mm x 9mm x 625 µm | SMA | Type 2 |
| Sapphire-G | 4.5mm x 4.5mm x 625 µm | SMA | Diamond-type |
| Sapphire-H | 9mm x 9mm x 625 µm | SHV | Sandwich |

*Table 2: List of the sapphire detector prototypes developed in this work, along with relevant characteristics.*

The detector bias was supplied by a CAEN NDT1470 NIM power module. Both positive and negative biases were tested, finding out that positive biases produced the most recognizable signals: as such, all the results in section 4 will refer to positive biases. Two amplification stages were tested to shape the signal: an Ortec 142A charge-sensitive preamplifier and a CAEN A422A charge-sensitive preamplifier, both coupled to an Ortec 672 Spectoscopic Amplifier. Despite the former is guaranteed only for biases up to ±1kV, the 142A setup demonstrated the best results for all the high biases used in section 4, and as such was adopted for all the measurements in that section.

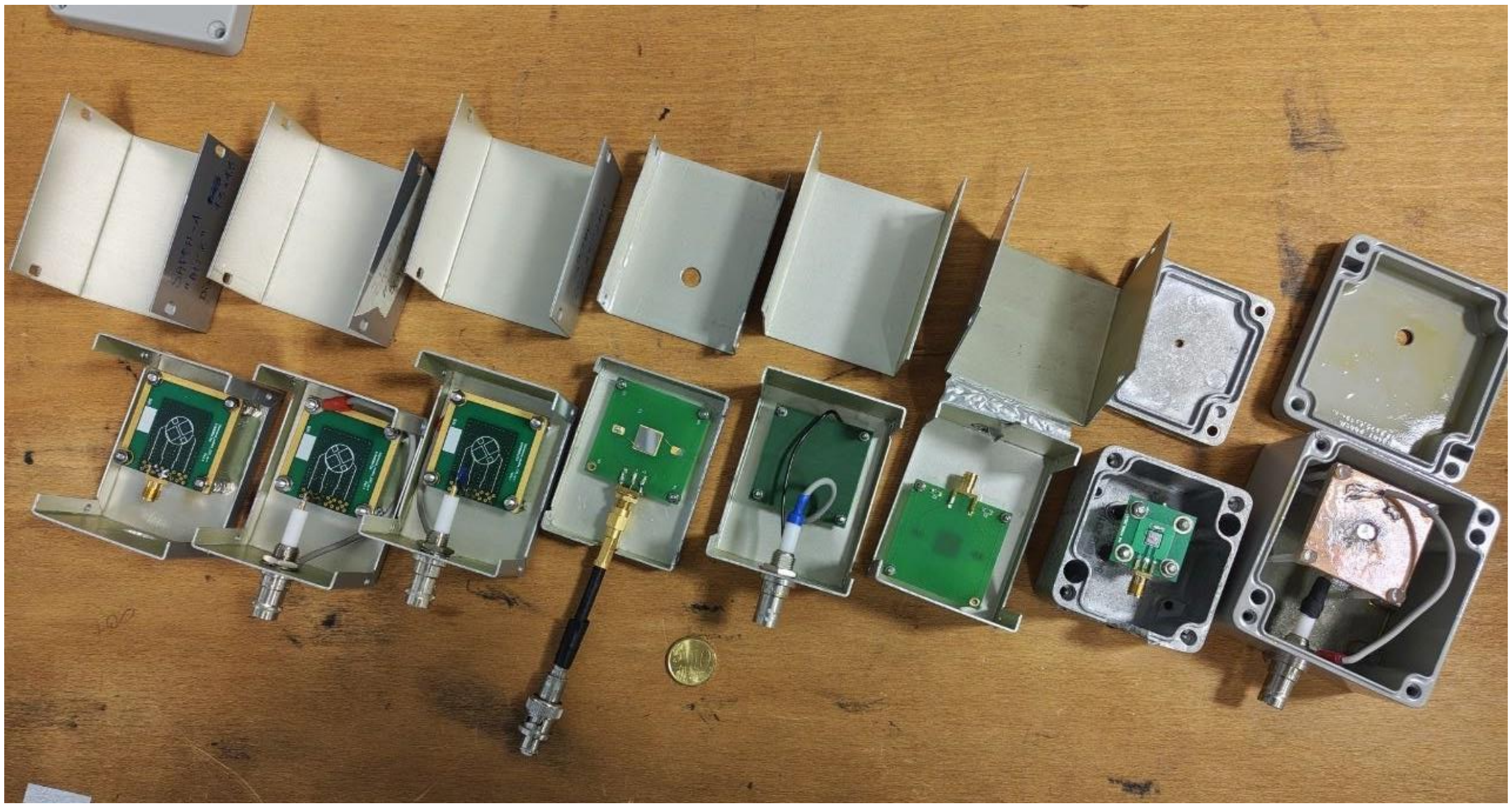

*Figure 4: Picture of the eight sapphire detector prototypes, lined up from Sapph-A (on the left) to Sapph-H (to the right). A coin is also featured for a reference of scale.*

## 4. Alpha Particle Irradiation and HV Scan

To test the functionality of the detectors, the prototypes connected to the electronic chain (see section 3) were irradiated with 5.47 MeV alpha particles emitted by an electrodeposited $^{241}$Am source. In order to minimize the energy loss in air, the source was kept as close as possible to the detector surface (between 5 and 15 mm, depending on the prototype's casing): as such, the alpha particle interacting with the prototypes had constant energy (for each detector) between 4 MeV and 5 MeV (35).

Each of the prototypes featured recognizable signals under irradiation, albeit small (≃30/40mV signal over a typical peak-to-peak noise of ≃8mV). The prototypes also underwent an HV scan to study the behaviour of such signals with changing bias. Signals were sampled with a 500MHz CAEN 5730 analog-to-digital converter. Pulse height (PH) spectra were produced, comparing the PH distribution of the same prototype with different biases. The HV scan spanned from the lowest bias values for which alpha particle detection was achieved (usually 1kV) to the highest bias value not featuring microdischarges, identified as non-alpha related signals with a non-negligible rate compared with alpha signals (usually between 2kV and 2.5kV). An exemplificative result is shown in Figure 5.

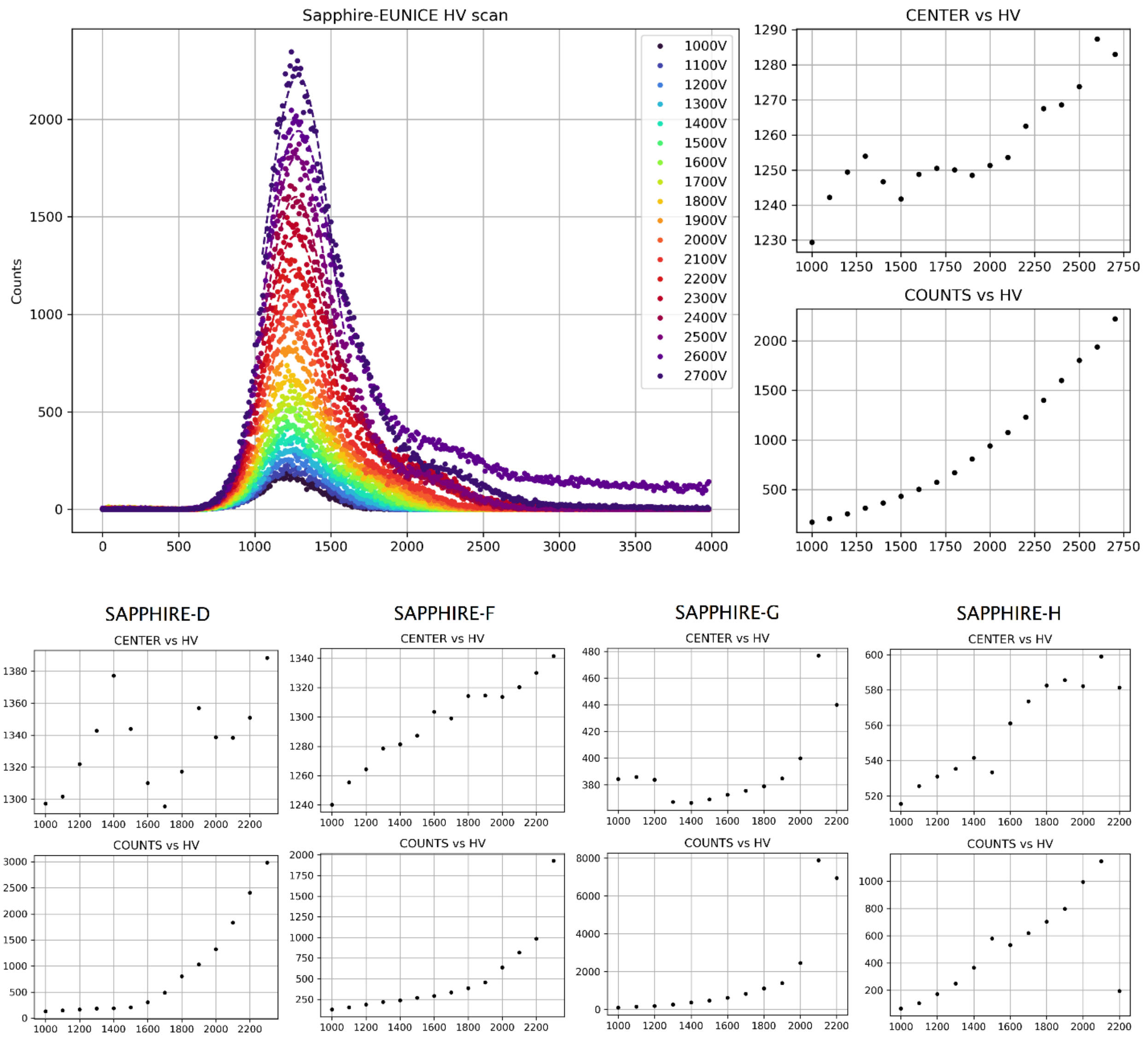

*Figure 5: On top: Pulse height spectra obtained with a sapphire prototype during a HV bias scan. The graphs on the right represent the x-axis location ("centre") of a gaussian fit of the peak and the number of counts in a 2-minutes time as a function of the bias. On bottom: centre and number of counts evolution for other 4 sapphire prototypes.*

The HV scan reveals that a bias increase also increases monotonically the number of counts per unit time of the detector. This is undesirable for a working detector, pointing to the fact that sapphires should either be coupled to biases higher than the ones measured in this work (up to 2.7kV) or thinner active volumes should be used. Increasing the bias also increases the position of the peak in the PH spectrum, possibly because the higher bias increases the charge collection efficiency of the prototype; despite this, the effect is much smaller than the increase in the number of counts. Results are fairly consistent between smaller (4.5x4.5 mm²) and larger (9x9 mm²) detectors, demonstrating that an increase in the size of the active volume does not negatively influence detection quality: this will be an important support for the test of larger devices (18x18 mm² and 36x36 mm²) whose sizes will be much larger than the available SCDs. Surprisingly, the experimentation revealed that the prototypes using SMA connectors wielded slightly better results in terms of signal-to-noise ratio, despite SMA standard being certified only up to ±500V. This might be due to the fact that the SMA standard is designed as a signal-carrier, while SHV standard is usually designed as a power-carrier – thus, the quality of the signal is better preserved on the former. Further tests will be necessary to corroborate the results.

## 5. Conclusions and Perspectives

This work presented the first steps towards the use of synthetic sapphire as the active material of solid-state detectors for fast neutrons. Eight prototypes with different active volume sizes, connectors and PCB designs were manufactured from commercial sapphire wafers and characterized with a $^{241}$Am alpha source. Geant4 simulations were conducted to estimate the expected response to DT neutrons.

All prototypes successfully detected alpha particles. This shows not only that sapphire can operate as an active detection SSD material, but also that the fabrication process developed for these prototypes, from metallization and dicing to mounting and packaging, yields working devices. Results were also consistent between 4.5 × 4.5 mm² and 9 × 9 mm² devices, supporting the scale-up to larger active areas. The signal, however, remains small, with typical amplitudes of 30–40 mV over a peak-to-peak noise of about 8 mV. Moreover, the count rate increased monotonically with bias over the whole range explored, without reaching a plateau where it becomes independent of the applied voltage. This indicates that sapphire detectors require higher electric fields than those used in this work, either pointing toward setups using higher biases or new prototypes with thinner active volumes.

A second generation of prototypes is currently being manufactured with the aim of reaching higher biases. In particular, a 500 nm thick oxide passivation layer is being added on the front electrode to prevent the microdischarges from the capacitor borders, which limited the bias to about 2.5 kV in this work. The design includes also very large bonding areas which will accommodate thicker bonding wires, minimizing the field in the wires and the probability of discharges. The second generation will also help determine the required operating bias, identify ways to improve the signal-to-noise ratio and define a more mature detector design. Once this design is consolidated, the detectors will be tested in a DT neutron field and their response compared with the Geant4 simulations.